\documentclass[twocolumn, pre,floatfix]{revtex4}
\usepackage{eurosym}
\usepackage{graphicx,amssymb,amstext,amsmath}
\usepackage{epsfig}
\usepackage[colorlinks=true, citecolor=blue, linkcolor=blue, urlcolor=blue, anchorcolor=blue]{hyperref}
\usepackage{epstopdf}
\usepackage[english]{babel}
\usepackage[autostyle]{csquotes}
\usepackage{lipsum}
\usepackage{subfigure,url}
\usepackage{color,graphics}
\usepackage{braket}
\usepackage{float}
\usepackage{mathtools}
\usepackage[pass,paperwidth=8.5in in,paperheight=11in]{geometry}
\draft
\newcommand*{\addheight}[2][.5ex]{%
  \raisebox{0pt}[\dimexpr\height+(#1)\relax]{#2}%
}
\begin{document}

\title{One-Dimensional Quantum Walks with a Position-Dependent Coin} 

\author{Rashid Ahmad, Uzma Sajjad, Muhammad Sajid}

\affiliation{Department of Physics, Kohat University of Science and Technology, Kohat 26000, Khyber-Pakhtunkhwa, Pakistan}

\date{\today}
\begin{abstract}
We investigate the evolution of a discrete-time one-dimensional quantum walk driven by a position-dependent coin. The rotation angle which depends upon the position of a quantum particle parameterizes the coin operator. For different values of the rotation angle, we observe that such a coin leads to a variety of probability distributions, e.g. localized, periodic, classical-like, semi-classical-like, and quantum-like. Further, we study the Shannon entropy associated with position space and coin space of a quantum particle and compare it with the case of the position-independent coin. We show that the entropy is smaller for most values of the rotation angle as compared to the case of the position-independent coin. We also study the effect of entanglement on the behavior of probability distribution and Shannon entropy of a quantum walk by considering two identical position-dependent entangled coins. We observe that in general, a quantum particle becomes more localized as compared to the case of the position-independent coin and hence the corresponding Shannon entropy is minimum. Our results show that position-dependent coin can be used as a controlling tool of quantum walks.
\end{abstract}
\keywords{}

\maketitle 


\section{Introduction}
Classical random walks are widely-used fundamental models of natural sciences in which the position of a classical particle is shifted depending upon the outcome of a classical coin. They have found applications in different research areas including diffusion and mobility in materials \cite{diffusion}, Brownian motion \cite{brownian}, polymer statistics \cite{slutsky}, exchange rate forecast \cite{exchange}, solution of differential equations \cite{diff}, quantum Monte Carlo techniques \cite{monte carlo}, and randomized algorithms \cite{algo}. Quantum random walks, motivated by classical random walks, were initiated by Aharonov et al. in 1993 \cite{aharonov}. In quantum walks, a unitary process for a quantum particle (also known as a walker), i.e. electron, atom or photon, replaces the stochastic evolution of a classical particle. Thus quantum walks become fundamentally different from its classical counterpart due to quantum interference that arises when different trajectories of the walker intersect each other hence leads to a very different probability distribution \cite{kempe, andracas}.
\par
There are basically two kinds of quantum walks the properties of which have already received a good deal of attention: continuous quantum walks \cite{cont1,cont2,cont3} and discrete-time quantum walks (DTQWs) \cite{disc1,disc2,disc3}. The latter of the two has received much attention because of their application in various disciplines. In physics, DTQWs are employed for the simulation of different physical phenomena, e.g. topological phenomena in condensed matter systems \cite{topo1,topo2,topo3,chiral}, quantum percolation \cite{perlocation}, and in evaluating the impact of the disorder on the dynamics of a walker \cite{disorder1,disorder2,disorder3}. In quantum information sciences, due to the faster expansion of quantum walks compared to the classical ones, DTQWs are used to develop fast algorithms for computations on quantum computers  \cite{search1,search2,speedup,algorithm1,algorithm2,algorithm3,algorithm4}, and to engineer and control certain quantum computation tasks \cite{computation1,computation2,computation3}.
\par
These applications have motivated a study of various properties of DTQWs \cite{dim,graphs,mcoin,mparticle,aperiodic,decoherent,history,exp3,exp4} and have been realized in numerous experiments using different physical settings including atoms \cite{karski,genske,robens}, photons \cite{exp1,broome,schreiber,sansoni,exp5}, trapped ions \cite{zahringer,exp2}, and superconducting qubits \cite{flurin}. To control their behavior, different types of walks have been studied, e.g. DTQWs with decoherence \cite{Kendon,albertinjp}. In a similar vein, recently coherent quantum walks with a step-dependent coin (SDC) has been studied \cite{panahiyan}. Such a coin can be used to get the desired probability distribution by controlling the rotation angle of the so-called coin operator. The authors considered different rotation angles and observed diverse probability distributions which they classified into different classes. This becomes a motivation for us that quantum walks can also be controlled by using a position-dependent coin (PDC).
\par
Moreover, quantum walks of a group of interacting particles in the realm of quantum computing \cite{quantcomp} made it possible to link the phenomenon of quantum entanglement \cite{e} with it. The entanglement of quantum bits will allow quantum computers to perform certain calculations much faster than the common classical computers. In this context progress has been made, e.g. quantum walks on a line with two entangled particles \cite{eparticles}, entanglement and interaction in a topological quantum walk \cite{etopo}, entanglement in coined quantum walks on regular graph \cite{egraph} etc. In addition, a mathematical model was formulated to understand unrestricted quantum walks using two entangled coins \cite{ecoins}. They demonstrated that the behavior of the walk with entangled coins is very different as compared to the usual quantum walks using a single coin. Moreover, S. Panahiyan and S. Fritzsche reported on one-dimensional quantum walks with four internal degrees of freedom, i.e. two entangled coins \cite{panahiyane}. They have shown that entanglement between two coins could be used as a resource for obtaining different probability distributions in position space. We extend this study to one-dimensional quantum walks with two position-dependent entangled coins (PDEC).

This article is organized in the following way: In sec. \ref{position dependent}, we outline the mathematical framework of the PDC and explain the classification of probability distributions. We discuss the Shannon entropy of the quantum walk with PDC and compare it to the case of the position-independent coin (PIC). We also show the standard deviation for the different classes of the probability distributions. In sec. \ref{epos}, we study quantum walks with PDEC in 1D. We compute the entropy in this case and compare it with position-independent entangled coins (PIEC). We conclude and give a brief outlook of our work in sec. \ref{conc}.

\section{Quantum Walk with PDC}\label{position dependent}
In a DTQW the evolution of a walker is controlled by a couple of unitary operators including a \enquote{coin operator} and a \enquote{conditional shift operator}, applied repeatedly. A coin operator acts on the internal state of a walker, which for a walker with two internal degrees of freedom (similar to the spin-1/2 particle) are generally known as spin-up state and spin-down state. The conditional shift operator acts on the external degree of freedom and moves a walker either to the left or right depending on its internal state. Here we consider a single-qubit walker, i.e. a walker with two internal degrees of freedom on a one-dimensional lattice. For a single-qubit walker, the coin (Hilbert) space, $\mathcal{H}_c$, is spanned by ${\ket 0, \ket 1}$. The position (Hilbert) space, $\mathcal{H}_p$, is spanned by ${\ket x: x \epsilon\:\mathbb{Z}}$. Thus the Hilbert space of the system is the tensor product of Hilbert spaces of the components, i.e. $\mathcal{H}_p \otimes \mathcal{H}_c$. The walker's internal degrees of freedom play an important role to have many features of DTQWs. The PDC operator is defined as

\begin{align}
\nonumber \hat{C}=&(\cos \hat{x}\theta \:  \ket 0_C \bra 0 +\sin  \hat{x}\theta\: \ket 0_C \bra 1
\label{coin} \\+&\sin \hat{x}\theta \: \ket 1_C \bra 0 -\cos  \hat{x}\theta\: \ket 1_C \bra 1)\otimes \ket x_P \bra x,
\end{align}
here $\theta$ is the rotation angle and $\hat{x}$ is the position operator acts on the external degree of freedom.  In contrast to previously studied inhomogeneous quantum walks \cite{algorithm3, AKempe,inhomo1, inhomo2}, we characterize the coin by a single rotation angle which depends on the position of the walker. Thus choosing different rotation angles results in a diverse probability distributions. The conditional shift operator is defined as

\begin{align}\notag
\hat{S}=&\ket 0_C \bra 0 \otimes \sum_x \ket {x+1}_P \bra x
\label{shift} \\+&\ket 1_C \bra 1 \otimes \sum_x \ket {x-1}_P \bra x.
\end{align}
This shift operator shifts the walker in spin-up state to the right and the one in spin down state to the left by one lattice site. The evolution operator (or the walk operator), $\hat{U}=\hat{S}\hat{C}$, acts on the initial state ($\phi_{\text{int}}$) of the walker within the product space $\mathcal{H}$. The repeated application of the walk operator to the initial state for large number of steps results in the evolution of the walk, i.e.

\begin{align}
\ket \phi_T=\hat{U}^T\ket \phi_{\text{int}},
\end{align}
here $\phi_{\text{T}}$ is the final state after $T$ steps of the walks.
\par
We consider the following initial states to study the evolution of the quantum walk with position dependent coin operator

\begin{align} \label{initial1}
 \ket \phi_{\text{int1}}= &\ket 0_C \otimes \ket 0_P,
 \\   \ket \phi_{\text{int2}}= &\ket 1_C \otimes \ket 0_P.
\end{align}
The states $\phi_{\text{int1}}(\phi_{\text{int2}})$ describes a particle prepared in spin up (spin down) internal state which is spatially localized around the origin of the 1D lattice. The stepwise evolution of the initial state $\phi_{\text{int1}}$ of the walker can be mathematically written as,

\begin{align}
\notag \ket 0_C \otimes \ket 0_P &\xRightarrow[\text{}]{\text{1st}}  \ket 0_C\otimes \ket 1_P,
\\ \notag &\xRightarrow[\text{}]{\text{2nd}}\cos \theta \: \ket 0_C \otimes\ket {2}_P+\sin \theta \:  \ket 1_C\otimes \ket 0_P,
\\\notag &\xRightarrow[\text{}]{\text{3rd}}\cos2 \theta \cos \theta\: \ket 0_C \otimes\ket {3}_P+\sin2 \theta \cos \theta
\\\notag & \ket 1_C \otimes\ket {1}_P-\sin \theta \:  \ket 1_C\otimes \ket {-1}_P,
\\ \label{f1} \xRightarrow[\text{}]{\text{Tth}} &\Bigg( \prod^{T-1}_{n=0} \cos n \theta  \Bigg)  \: \ket 0_C \otimes\ket {T}_P+.......,
\end{align}
and similarly for the second choice of the initial state, we have
\begin{align}
\notag \ket 1_C \otimes \ket 0_P &\xRightarrow[\text{}]{\text{1st}} - \ket 1_C\otimes \ket {-1}_P,
\\ \notag &\xRightarrow[\text{}]{\text{2nd}}\sin \theta \: \ket 0_C \otimes\ket {0}_P+\cos \theta \:  \ket 1_C\otimes \ket {-2}_P,
\\\notag &\xRightarrow[\text{}]{\text{3rd}}-\sin2 \theta \cos \theta\: \ket 0_C \otimes\ket {-1}_P+\sin \theta
\\\notag &\ket 0_C \otimes\ket {1}_P-\cos2 \theta \cos \theta \:  \ket 1_C\otimes \ket {-3}_P,
\\ \label{f2} \xRightarrow[\text{}]{\text{Tth}}&(-1)^{T} \Bigg( \prod^{T-1}_{n=0} \cos n \theta  \Bigg)  \: \ket 0_C \otimes\ket {-T}_P+......
\end{align}
From the final states of the walker after certain number of steps T, it is found that a walker can occupy at most T number of lattice sites. The final states also show that due to quantum interference a walker initially prepared in state $\phi_{\text{int}1} (\phi_{\text{int}2})$ has zero probability at x=-T (x=T) position. Also for both initial states ($\phi_{\text{int1}}$ and  $\phi_{\text{int2}}$), the occupied positions are always even for even number of steps and for odd number of steps the occupied positions are odd.

\subsection{Classification of Probability Distributions}\label{probability density}
From Eq. (\ref{f1}) and (\ref{f2}) it is clear that the evolution of the quantum walk depends on the rotation angle ($\theta$). In this section, we make use of this fact to feature the possible classes of the probability distribution of this walk by choosing different rotation angles. We carry out numerical simulations for the initial state $\phi_{\text{int1}}$ and classify the behavior of probability distributions of the walk into the following classes.
\newline
\textbf{Free localized walk}: For rotation angle $\theta=0$, the evolution of the walk shows that the walker is at position $x=T$ with probability 1 after T steps as shown in Fig. \ref{fig:a}a. For this rotation angle, the coin operator is the identity operator and hence does not change the internal state of the walker. As a result, the walker initially prepared in the spin-up state remains in the same internal state during the evolution of the walk and shifts to right by a unit distance at each step of the walk.
\newline
\textbf{Bounded localized walk}: For rotation angle $\pi/2$, the probability of finding the walker is bounded to at most three positions ($x=0,\pm1$) irrespective of the number of steps. The walker is localized because the probability is 1 at any of the three possible positions. The probability distribution of the walker after T=30 steps of the walk is shown in Fig. \ref{fig:a}b. This localization can be explained as follows. The coin operator is identity operator at the initial position $x=0$ and acts as a spin flip operator (changing spin up to spin down and vice versa) at $x=\pm1$. Hence a walker initially prepared in the spin-up state will shift to the right during its evolution and will be reflected back from $x=1$. It will start moving to left but will be reflected back again from the $x=-1$ position. This way the walker will oscillate between $x=+1$ and $x=-1$. For this rotation angle, the walk with PIC shows that the walker is localized and bounded to two positions ($x=-1,0$).
\newline
\textbf{Bounded Periodic walk}: For rotation angle $\pi/4$, the probability distribution shows a periodic behavior of localized, two-peaks-zone and three-peaks-zone for a different number of steps. The probability is 1 at $x=0$ position (localized) for certain number of steps, or equally distributed between positions $x=\pm1$ (two-peaks-zone), or distributed among positions $0,\pm2$ with probabilities 0.5, 0.25 and 0.25 respectively (three-peaks-zone). The probability distribution is shown in Fig. \ref{fig:a}c for T=30 which shows two-peaks-zone. Moreover, for this rotation angle, the coin operator becomes a spin flip operator at $x=\pm2$ and hence the walker reflects back from both sides and remains trapped between these positions. For this rotation angle, the PIC is a simple Hadamard coin which shows asymmetric probability distribution.
\newline
\textbf{Bounded classical-like walk}: For rotation angle $\pi/6$, the probability distribution shows Gaussian-like behavior, i.e. maximum probability of finding the walker is maximum at (or near) $x=0$ position and there is comparatively smaller but non zero probability of finding the walker around the most probable position (see Fig. \ref{fig:a}d). Moreover, the walker is bounded to 7 lattice sites and the coin operator becomes a spin flip operator at $x=\pm3$ and hence the walker gets reflected from these positions. For this rotation angle, a walk with PIC is not bounded and the walker spreads away from its initial position.
\newline
\textbf{Classical-like walk}: For rotation angle $7\pi/45$, the probability distribution shows classical-like behavior but is not bonded. For this rotation angle, a walk with PIC shows the maximum probability of finding the walker towards the sides of the distribution (Fig. \ref{fig:a}e).
\newline
\textbf{Fast classical-like walk}: For rotation angle $\pi/5$, the probability distribution of the walk is Gaussian-like but the number of positions occupied by the walker is greater compared to the classical-like walk. Moreover, in the case of the fast classical-like walk, the number of occupied positions increases more rapidly than the case of the classical-like walk. For this rotation angle, PIC shows maximum probability towards the side (Fig. \ref{fig:a}f).
\newline
\textbf{Semi-classical-like walk}: For rotation angle $\pi/3$, the probability of finding the walker is maximum near the center of occupied positions. There are other significant peaks towards sides of the highest peak. We call this walk as the semi-classical-like walk. For this rotation angle, a walk with PIC shows maximum probability towards the right (Fig. \ref{fig:a}g).
\newline
\textbf{Bounded Semi-classical-like walk}: For rotation angle $\pi/30$, the probability distribution is semi-classical-like but is bounded to 31 lattice sites. The coin operator becomes a spin-flip operator at $x=\pm15$ and thus the walker cannot surpass these positions. We call the walk with this type of probability distribution as the bounded semi-classical-like walk. The PIC for this rotation angle shows maximum probability towards the right (Fig. \ref{fig:a}h).
\newline
\textbf{Fast Semi-classical-like walk}: For rotation angle $\pi/3$, the probability distribution is semi-classical-like but the number of occupied positions is greater compared to the semi-classical-like walk. The PIC for this angle shows semi-classical-like but not fast. (Fig. \ref{fig:a}h).
\newline
\textbf{Bounded quantum-like walk}: For rotation angle $\pi/90$, the walk shows ballistic, asymmetric expansion. The walk is quantum-like but is bounded to finite positions. For this rotation angle, the coin operator becomes a spin-flip operator at $x=\pm45$. We call this bounded quantum-like walk. For this rotation angle, PIC shows maximum probability at position $x=T$ (Fig. \ref{fig:a}j).

\begin{center}
\begin{table}
\begin{tabular}{p{6cm}p{2cm}}
\hline
\textbf{Classes of the walk} \par& \textbf{Rotation angle} ($\mathbf{\theta}$) \\
\hline
Free localized walk \par&   \quad 0\\

Bounded localized walk \par & \quad $\pi/2$\\

Periodic walk \par & \quad $\pi/4$\\

Bounded classical-like walk \par & \quad $\pi/6$\\

Classical-like walk \par & \quad $7\pi/45$\\

Fast classical-like walk \par & \quad $\pi/5$\\

Semi-classical-like walk \par & \quad $2\pi/5$\\

Bounded semi-classical-like walk \par & \quad $\pi/20$\\

Fast semi-classical-like walk \par & \quad $\pi/3$\\

Bounded quantum-like walk \par & \quad $\pi/90$\\
\hline
\end{tabular}
\caption{Classification of the probability
distribution for quantum walks with position-dependent coin}
    \label{tab:a}
\end{table}
\end{center}

\subsection{Shannon Entropy} \label{sentropy}
The idea of Shannon entropy was presented by Claude Shannon in 1948 \cite{entropy}. In quantum walks, the Shannon entropy was introduced and numerically studied for various kinds of Hadamard coin \cite{ent1,ent2}. The Shannon entropy for each step of a walk is determined from the probabilities of positions and internal degrees of freedom occupied by a wave function. In quantum walks with PDC, the Shannon entropy becomes position-dependent. We compute the Shannon entropy associated with position space ($S_P$) and coin space ($S_C$) for the PDC and compare it with the case of PIC. The Shannon entropy is defined as

\begin{align}
&S_P= -\sum_x P_x \log_e P_x,
\\&S_C= -\sum_i P_i \log_e P_i,
\end{align}
here $P_x$ is the probability of finding the walker at position $x$ and $P_i$ is the probability of the walker with an internal degree of freedom $i$ where $i=0,1$. The entropies for different rotation angles in case of quantum walks with PDC and PIC are plotted in Fig. \ref{fig:b}. For a walk with PIC, the entropy of position space always increases with the number of steps (T), and this behavior does not change with the rotation angle. For a walk with PDC, we observe the following about the entropy of position space (we will equally call it entropy):
\par
\begin{itemize}
\item Entropy of a localized walk is always zero as there is no uncertainty in the state of the walker. This is the case for the two classes of the walk, i.e. free localized walk ($\theta= 0$) and bounded localized walk $(\theta = \pi/2)$.

\item For other rotation angles, the evolution of entropy shows three types of behaviors, i.e. bounded periodic, bounded, and unbounded behavior.

\item For a bounded periodic walk ($\theta=\pi/4$) the entropy is bounded periodic as the uncertainty in the state of a physical system periodically changes. Whenever the walker is localized the entropy of the state is $0$, for two-peaks-zone entropy is $0.7$ and for three-peaks-zone, the entropy is $1$ as in the Fig  \ref{fig:b}a.

\item Entropies of bounded walks (classical-like, semi-classical-like, and quantum-like walk observed at the rotation angles $\theta=\pi/6,\pi/20$, and $\pi/90$ respectively) are bounded but non-periodic. In this case, the entropies of the walk do not increase after reaching a certain maximum limit. The maximum limit is different for different rotation angles as shown in Fig. \ref{fig:b}(b),(e),(h).

\item For other classes of the walk the behavior of Shannon entropy is unbounded, i.e. it increases by increasing the number of steps.

\item We also found that the entropy for a quantum walk with PDC is small as compared to PIC except for the bounded quantum-like walk. The exception is because this class of the walk is observed for a small rotation angle (closer to 0). In this case, quantum walk with PIC shows localized-like behavior and hence the corresponding entropy of the state becomes smaller. The same trend in entropy is seen in the case of the bounded semi-classical-like walk but for a smaller number of steps of the walk. For semi-classical-like and fast semi-classical-like walks there is almost no difference in the entropies of the walks with PDC and PIC. In both types of walk, the entropy increases with the number of steps.
\end{itemize}
The Shannon entropy ($S_C$) associated with coin space of the walker in case of PIC shows periodic behavior irrespective of the value of the rotation angle. For a walk with PDC, we observe the following for $S_C$:
\begin{itemize}
\item For a periodic walk $S_C$ shows periodic behavior. When the walk is localized then $S_C$ is zero similar to $S_P$. This case takes place for specific values of T in a periodic walk.

\item For all the subclasses of the classical-like walks $S_C$ is small as compared to their PIC counterparts. Generally, our results show that the $S_C$ of a walk with PDC is small as compared to PIC except for the bounded semi-classical-like and bounded quantum-like walks. For semi-classical-like and fast semi-classical-like walks, $S_C$ is almost the same and increases with the number of steps in case of both PDC and PIC.
\end{itemize}
\onecolumngrid

\begin{center}
\noindent
\begin{figure}[H]
\begin{minipage}{0.98\columnwidth}

\begin{tabular}{ccc}
      \addheight {\includegraphics[width=57mm]{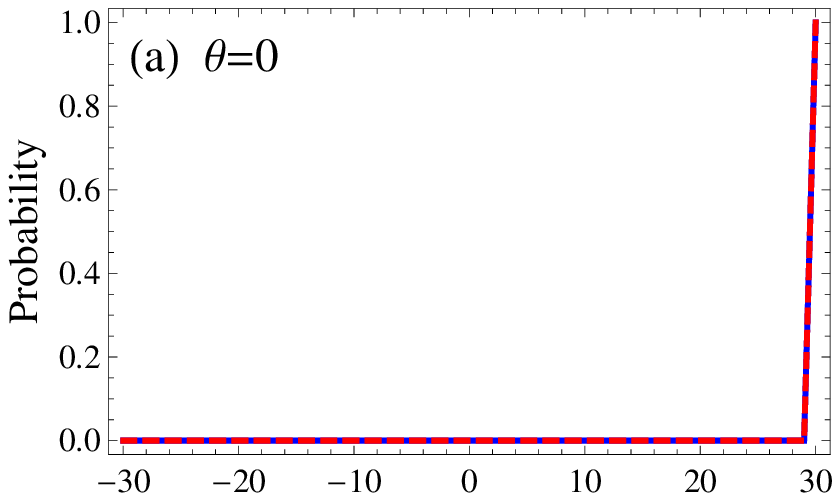}} &
      \addheight {\includegraphics[width=57mm]{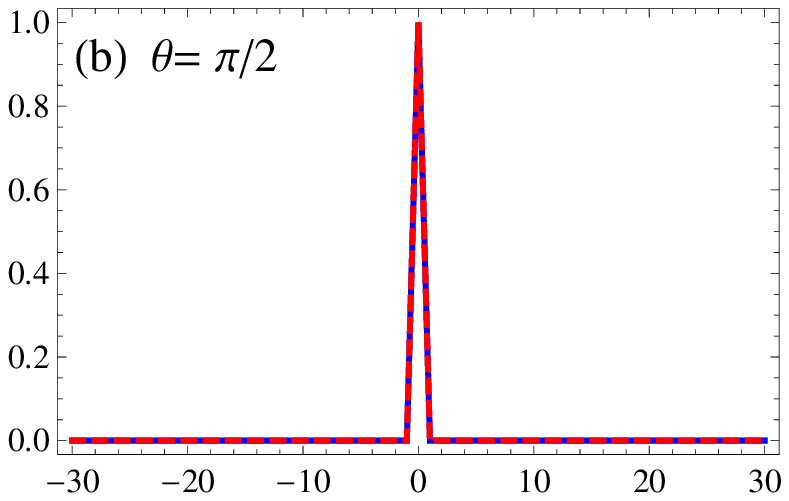}} &
      \addheight{\includegraphics[width=57mm]{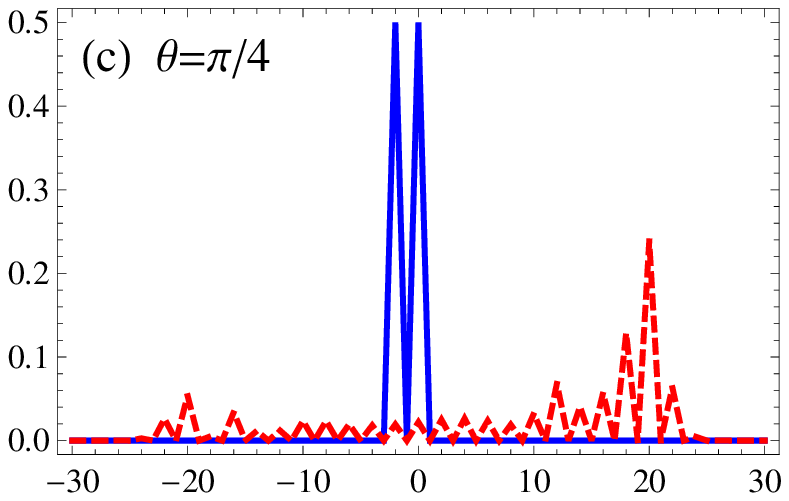}}\\
      \addheight{\includegraphics[width=57mm]{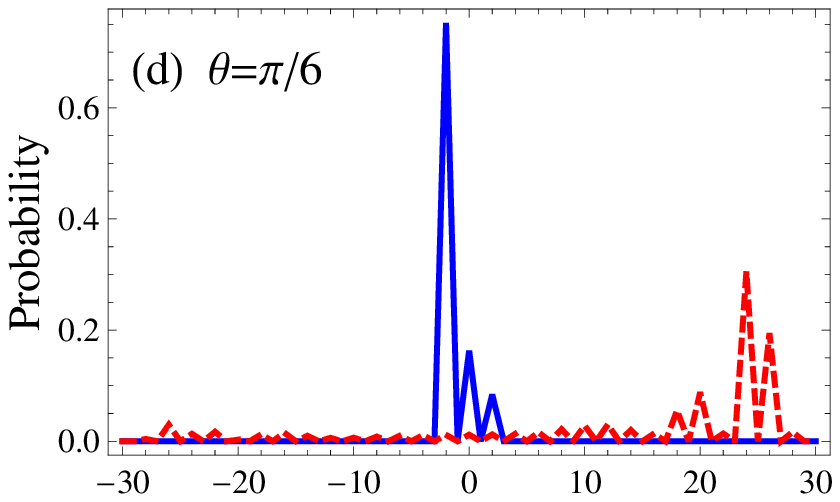}} &
      \addheight{\includegraphics[width=57mm]{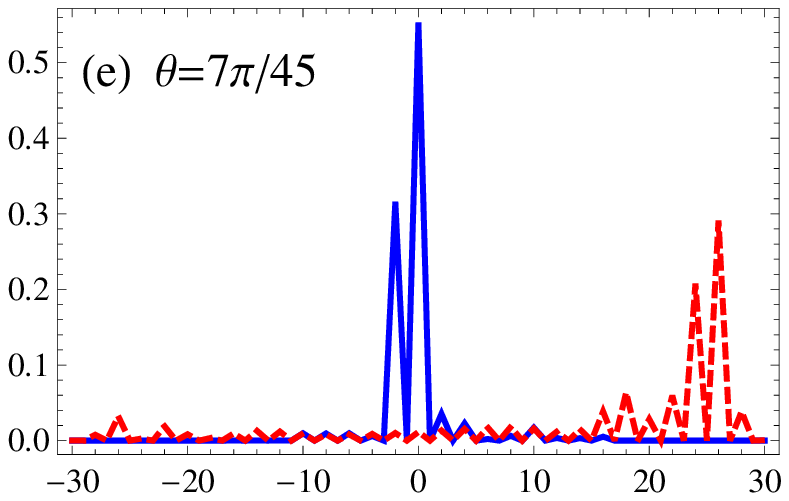}} &
      \addheight{\includegraphics[width=57mm]{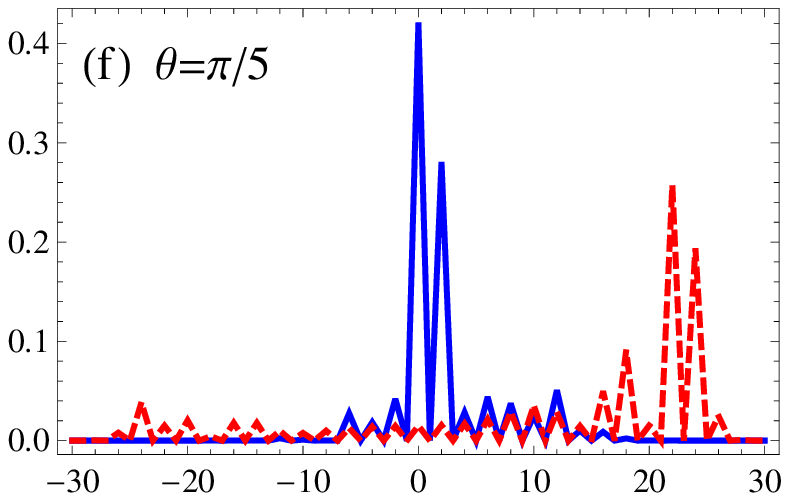}}\\
       \addheight{\includegraphics[width=57mm]{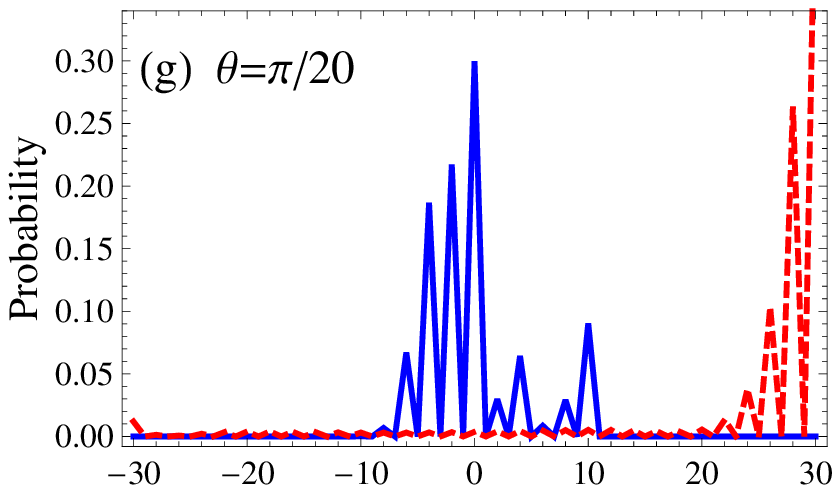}}&
        \addheight{\includegraphics[width=57mm]{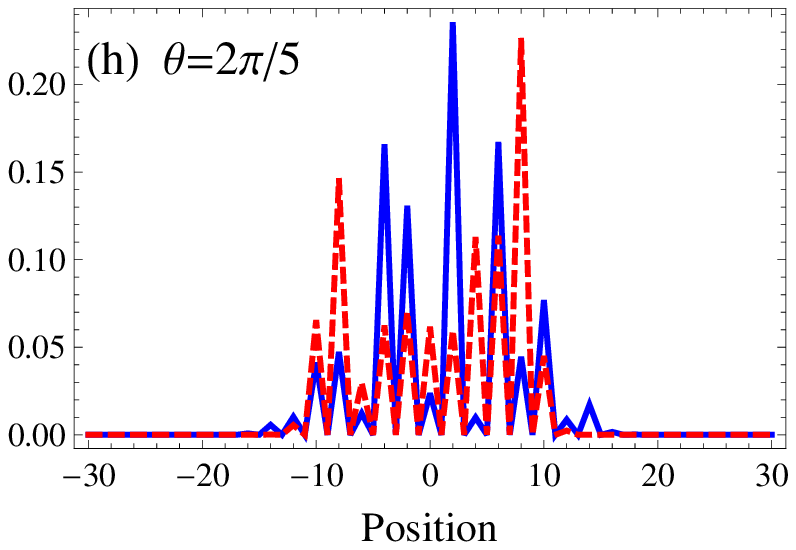}}&
        \addheight{\includegraphics[width=57mm]{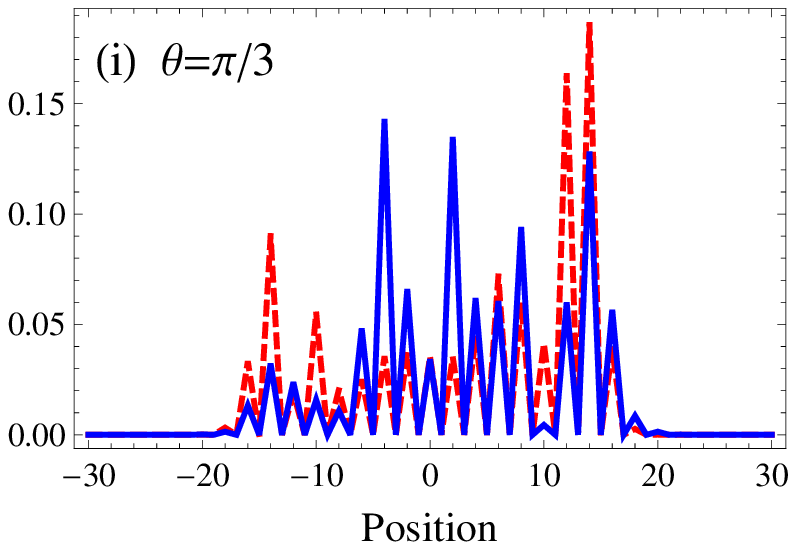}}\\
      \addheight {\includegraphics[width=57mm]{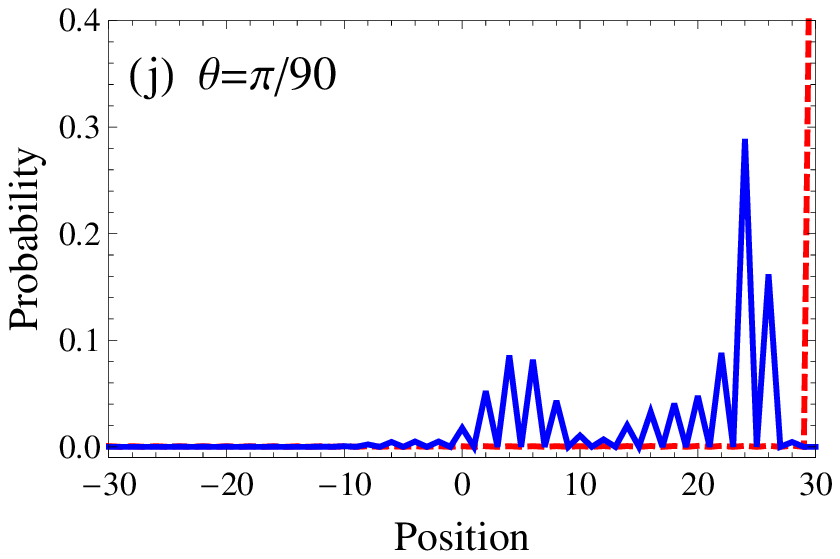}} \\
\end{tabular}
\caption{Probability distribution of the walk after T$=30$ for different rotation angles $\theta$ (indicated in the inset). Solid blue line represents the quantum walk with position-dependent coin and dashed red line represents the quantum walk with position-independent coin.}
\label{fig:a}
\end{minipage}
\end{figure}
\end{center}
\twocolumngrid

\onecolumngrid

\begin{center}
\noindent
\begin{figure}[H]
\begin{minipage}{0.98\columnwidth}

\begin{tabular}{ccc}
      \addheight{\includegraphics[width=57mm]{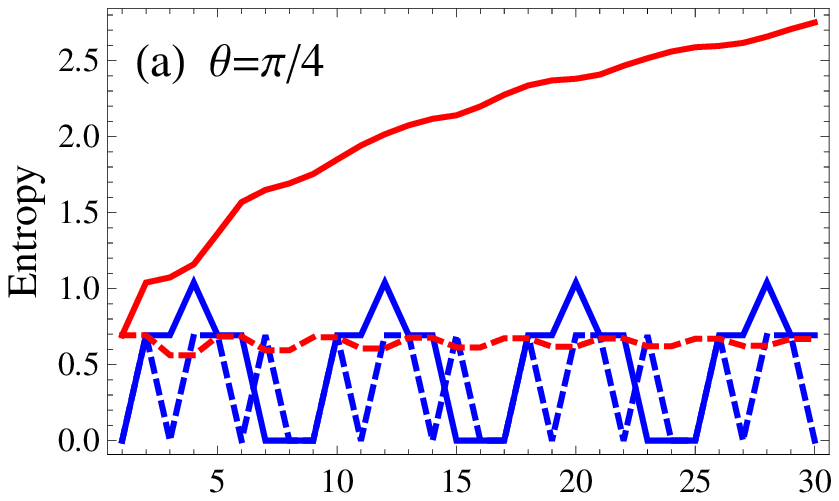}} &
     \addheight{\includegraphics[width=57mm]{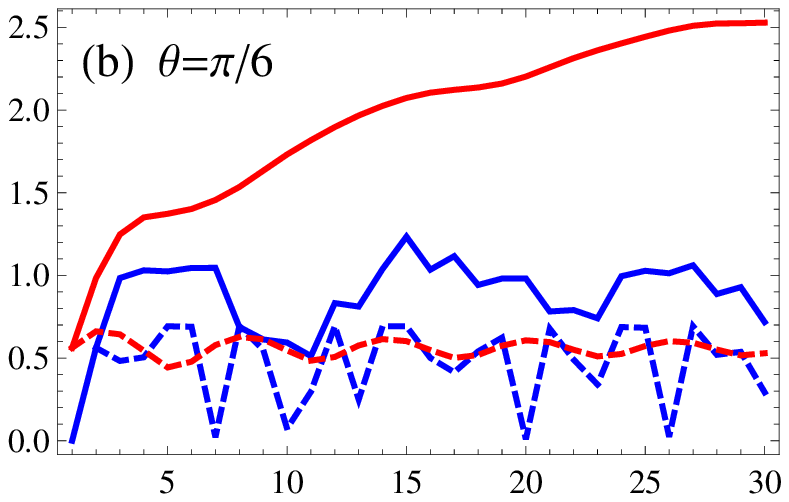}} &
      \addheight{\includegraphics[width=57mm]{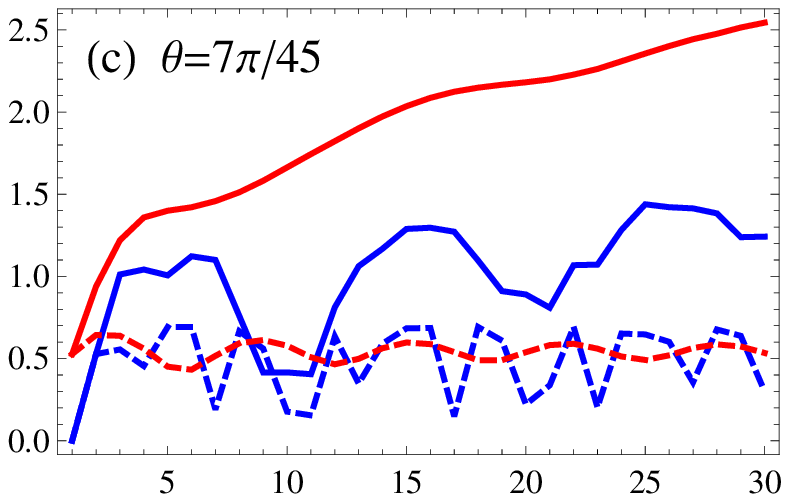}}\\
     \addheight{\includegraphics[width=57mm]{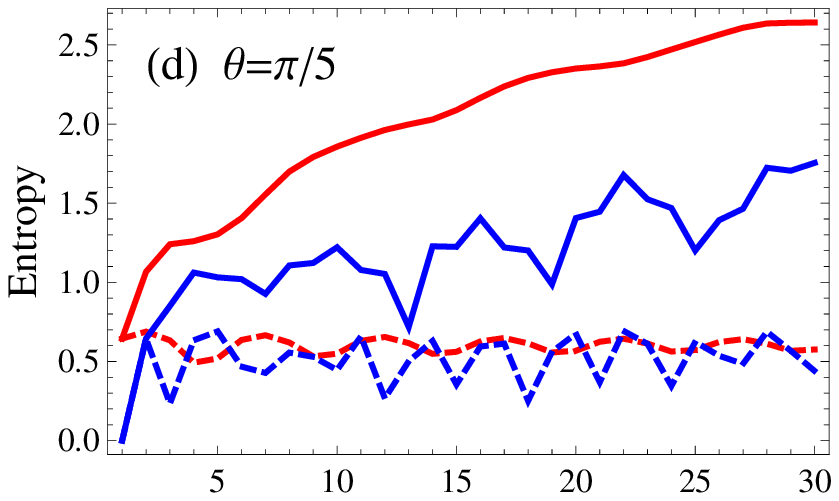}} &
      \addheight{\includegraphics[width=57mm]{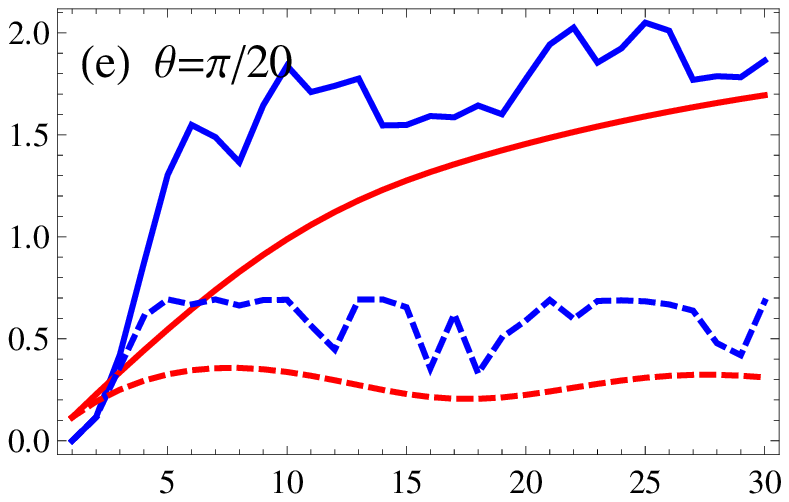}} &
      \addheight{\includegraphics[width=57mm]{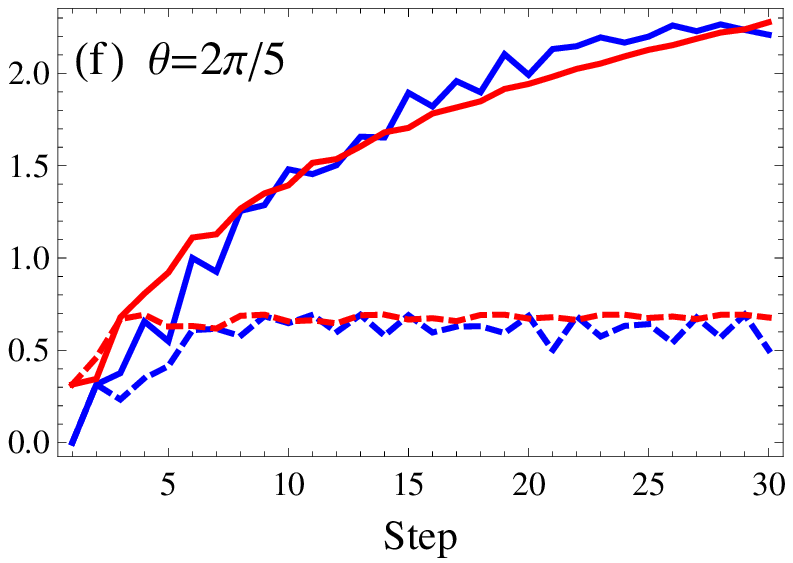}} \\
      \addheight{\includegraphics[width=57mm]{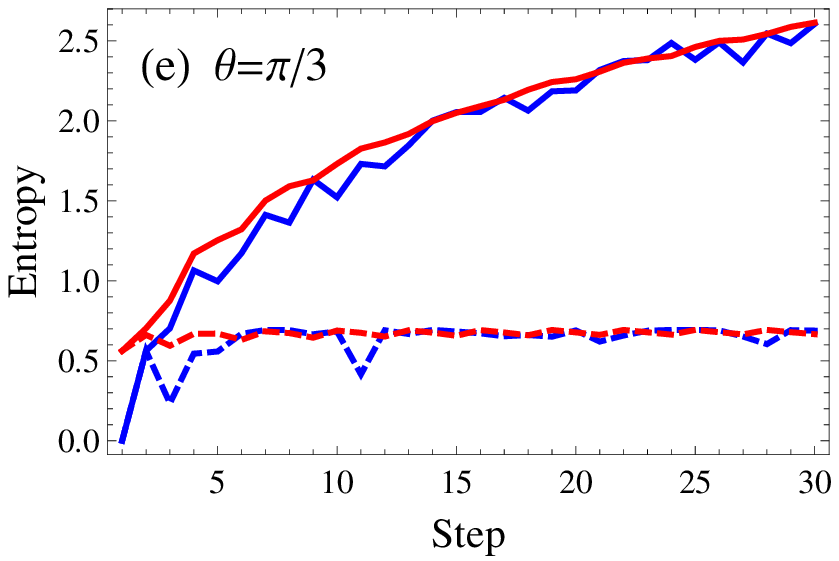}} &
     \addheight{\includegraphics[width=57mm]{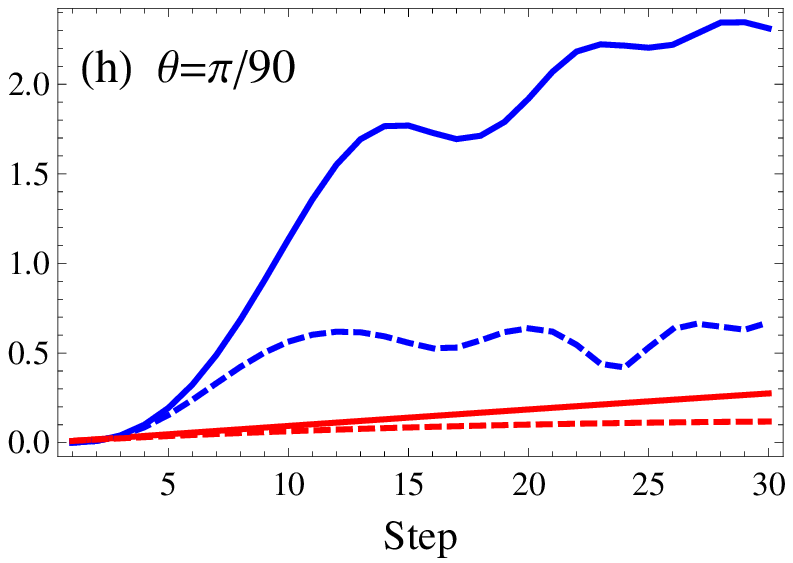}} \\
\end{tabular}
\caption{Shannon entropy vs the no of steps after T=30. Blue line represents the quantum walk with PDC and red line represents the quantum walk with PIC. Solid line shows the Shannon entropy associated to position space and dashed line shows the Shannon entropy associated to coin space.}
\label{fig:b}
\end{minipage}
\end{figure}
\end{center}
\twocolumngrid
\subsection{Standard Deviation} \label{sd}
We compute the standard deviation ($\sigma$) for all the classes of the quantum walk with PDC. The results are shown in Fig. \ref{fig:c3}. For a localized walk (both free and bounded) the standard deviation is 0 and for a periodic walk, it changes periodically. For other classes of the walk we observe the following:
\\*
(I) \textbf{Classical-like walk}
\\*
We compute the standard deviation of all the three subclasses of the classical-like walk and compare it with the $\sigma$ of the standard classical random walk (see Fig. \ref{fig:c3}a). The square root dependence of the $\sigma$ on the number of steps (T) of the walk justifies their classification as the classical-like walk. Moreover, for rotation angle, $\pi/5$ the $\sigma$ is greater than the other subclasses which show that for this rotation angle the classical-like walk spreads faster. In case of a bounded classical-like walk, the $\sigma$ does not increase from a certain maximum limit.
\\*
(II) \textbf{Semi-classical-like walk}:
\\*
For all the subclasses of the semi-classical-like walk, we compare the standard deviations of a quantum walk with position-independent Hadamard coin. The walker's initial state is chosen as in Eq. (\ref{initial1}). In case of normal and fast semi-classical-like walks, the standard deviations increase linearly with T similar to the case of simple Hadamard walk albeit rather slowly. This justifies the name of this class. In case of the bounded semi-classical-like walk, Initially the standard deviation increases linearly with T but this is not the case for higher T as the walker is bounded. This behavior is shown in Fig. \ref{fig:c3}b.
\\*
(III) \textbf{Quantum-like walk}:
\\*
For quantum-like walk we have only one subclass, i.e. bounded quantum-like walk. We compare the standard deviation of this walk to the standard deviation of a Hadamard walk. Our results show that the standard deviation of this walk increases linearly for initial steps, however, because of the fact the walker is bounded, the standard deviation does not increase further as shown in Fig. \ref{fig:c3}c.

\onecolumngrid

\begin{center}
\noindent
\begin{figure}[H]
\begin{minipage}{0.98\columnwidth}
\begin{tabular}{ccc}

      \addheight{\includegraphics[width=57mm]{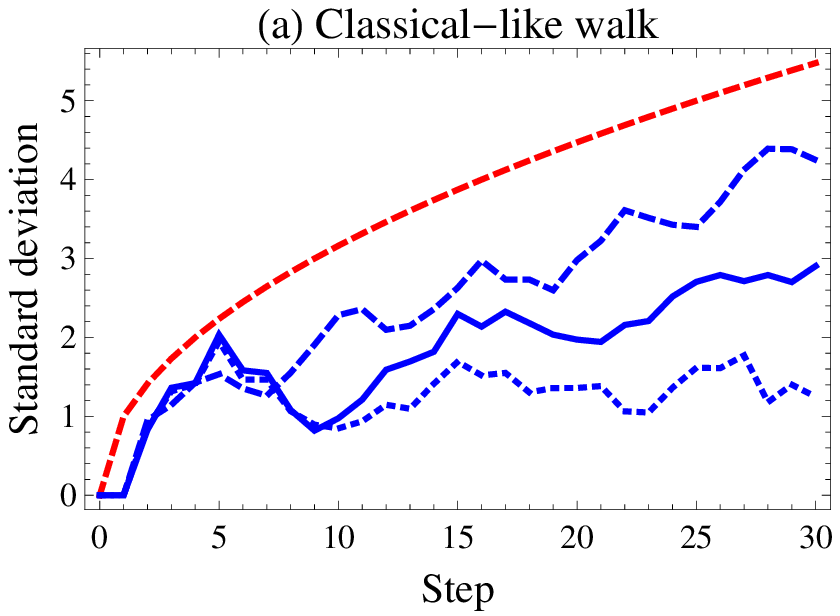}} &
     \addheight{\includegraphics[width=57mm]{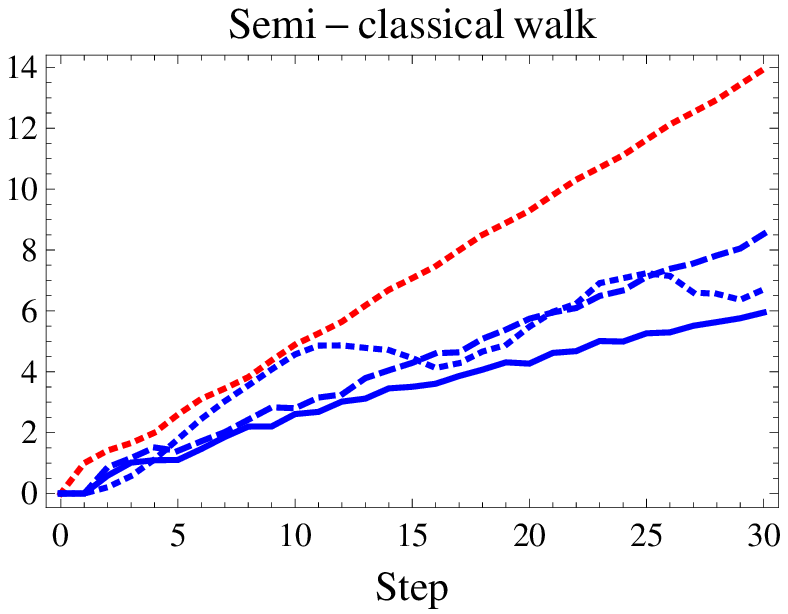}} &
      \addheight{\includegraphics[width=57mm]{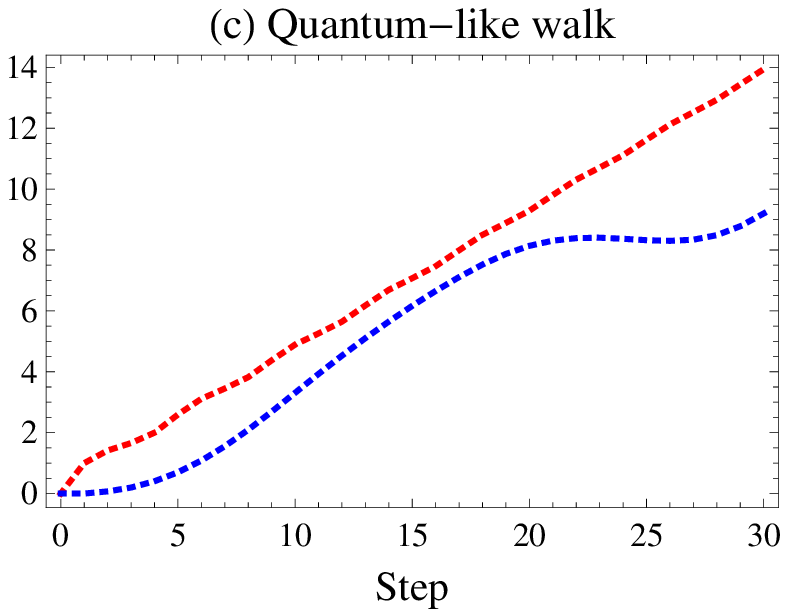}}\\
\end{tabular}
\caption{Standard deviation ($\sigma$) of the walk vs number of steps of the walk for T=30. Dashed red line shows the standard deviation of a simple classical random walk ($\sigma=\sqrt T$). Dotted red line shows the standard deviation of a simple Hadamard walk with position-independent coin ($\sigma \thickapprox T$). Blue lines show the standard deviation of the quantum walk with position-dependent coin. Dotted blue line represents bounded behavior of the walk. Solid blue line represents the normal walk (classical-like and semi-classical-like) and dashed blue line represents corresponding fast behavior of the walk.}
\label{fig:c3}
\end{minipage}
\end{figure}
\end{center}

\twocolumngrid

\section{Quantum walk with PDEC}\label{epos}
In the generalization of the quantum walk with PDC to PDEC the internal degrees of freedom of the walker become four. The initial state of the walker is maximally entangled. The Hilbert space $\mathcal{H}_{\text{EC}}$ associated with the internal states of the entangled two-qubit walker is spanned by ${\ket{00},\ket{01},\ket{10},\ket{11}}$. The entangled coin operator is given by the tensor product of the two identical PDCs given in Eq. (\ref{coin}), i.e. $\hat{\text{C}}_{\text{EC}}=\hat{\text{C}} \otimes \hat{\text{C}}$.

\begin{align}
\notag\hat{\text{C}}_{\text{EC}}=&(\cos^2 \hat{x}\theta \:  \ket {00}_{CC} \bra {00} +\sin  x\theta\cos \hat{x} \theta \: \ket {01}_{CC} \bra {00}
\\ \notag &+\sin  \hat{x}\theta\cos \hat{x} \theta \: \ket {10}_{CC} \bra {00} +\sin^2  \hat{x}\theta  \: \ket {11}_{CC} \bra {00}
\\ \notag&+\sin  \hat{x}\theta\cos \hat{x} \theta \: \ket {00}_{CC} \bra {01}-\cos^2 \hat{x}\theta \:  \ket {01}_{CC} \bra {01}
\\ \notag &+\sin^2 \hat{ x}\theta  \: \ket {10}_{CC} \bra {01}-\sin  \hat{x}\theta\cos \hat{x} \theta \: \ket {11}_{CC} \bra {01}
\\ \notag& +\sin  \hat{x}\theta\cos \hat{x} \theta \: \ket {00}_{CC} \bra {10}+\sin^2  \hat{x}\theta  \: \ket {01}_{CC} \bra {10}
\\ \notag &+\cos^2 \hat{x}\theta \:  \ket {10}_{CC} \bra {10}-\sin  \hat{x}\theta\cos \hat{x} \theta \: \ket {11}_{CC} \bra {10}
\\ \notag& +\sin^2  \hat{x}\theta \:  \ket {00}_{CC} \bra {11} -\sin  \hat{x}\theta\cos x \theta \: \ket {01}_{CC} \bra {11}
\\\notag &+\sin  \hat{x}\theta\cos x \theta \: \ket {10}_{CC} \bra {11} + \cos^2 \hat{x}\theta \: \ket {11}_{CC} \bra {11})
\\ \label{ecoinp}&\otimes \ket x_{P} \bra x.
\end{align}
The conditional shift operator in the case of PDEC is defined as
\begin{align}
\notag\hat{\text{S}}_{\text{EC}}=&\ket {00}_{CC} \bra {00} \otimes \sum\ket {x+1}_{P} \bra x
\\\notag &\ket {01}_{CC} \bra {01} \otimes \sum\ket {x}_{P} \bra x
\\\notag &\ket {10}_{CC} \bra {10} \otimes \sum\ket {x}_{P} \bra x
\label{eshift} \\+&\ket {11}_{CC} \bra {11} \otimes \sum \ket {x-1}_{P} \bra x.
\end{align}
The significant difference as compared to the quantum walk with PDC is that here the shift operator ($\hat{S}_{\text{EC}}$), in addition to the operations of shifting the walker to right and left, also behaves like an identity operator depending on the internal state of the walker. The coin's initial state can be chosen as

\begin{align}
\notag \ket \Phi_{\text{int}}=(\alpha_1 \ket {00}_{CC} +\alpha_2 \ket {01}_{CC} +\alpha_3 \ket {10}_{CC}
\\ \label{einitial}+\alpha_4 \ket {11}_{CC})\otimes \ket 0_{P},
\end{align}
where $\sum_{i}\alpha_i^2=1$  and the quantity $\alpha_i$ specifies the amount of entanglement. We consider two maximally entangled Bell states. (i) when $\alpha_2=\alpha_3=0$ and  $\alpha_1=\cos \eta$, $\alpha_4=\sin \eta$ where $\eta=\pi/4$. (ii) When $\alpha_1=\alpha_4=0$ and  $\alpha_2=\cos \eta$, $\alpha_3=\sin \eta$ with $\eta=\pi/4$. Thus the two maximally entangled initial states of the walker are
\begin{align}\label{int1}
\ket \Phi_{\text{int1}}= &\frac{1}{\sqrt{2}}(\ket {00}_{CC} + \ket {11}_{CC})\otimes \ket 0_P,
\\\label{int2}\ket \Phi_{\text{int2}}=&\frac{1}{\sqrt{2}}(\ket {01}_{CC} + \ket {10}_{CC})\otimes \ket 0_{P}.
\end{align}
Other choices of initial states are also possible but we select Eq. (\ref{int1}) and Eq. (\ref{int2}) to show the basic scheme of the maximally entangled state. For initial state $\Phi_{\text{int1}}$ the stepwise evolution of the walker, by applying the walk operator ($\hat{U}=\hat{S}_{EC}\hat{C}_{EC}$), can be mathematically written as,

\begin{align}\notag
&\frac{1}{\sqrt{2}}(\ket {00}_{CC} +\ket {11}_{CC})\otimes \ket 0_{P}
\\\notag&\xRightarrow[\text{}]{1\text{st}}\frac{1}{\sqrt{2}}(   \ket {00}_{CC} \otimes\ket {1}_{P}+ \ket {11}_{CC} \otimes\ket {-1}_{P}),
\\ \notag &\xRightarrow[\text{}]{2\text{nd}}\frac{1}{\sqrt{2}}(  \sin^2\theta \ket {00}_{CC} \otimes\ket {0}_P+ \cos^2 \theta \ket {00}_{CC} \otimes\ket {2}_{P}
\\ \notag &+\cos\theta\sin\theta \ket {01}_{CC} \otimes\ket {-1}_P+ \cos \theta\sin\theta \ket {01}_{CC} \otimes\ket {1}_{P}
\\\notag &+\cos\theta\sin\theta \ket {10}_{CC} \otimes\ket {-1}_{P}+ \cos \theta\sin\theta \ket {10}_{CC} \otimes\ket {1}_{P}
\\\notag &+\sin^2\theta \ket {11}_{CC} \otimes\ket {0}_{P}+ \cos^2 \theta \ket {11}_{CC} \otimes\ket {-2}_{P}),
\\ \notag &\xRightarrow[\text{}]{\text{Tth}}
\prod^{T-1}_{n=0}\frac{\cos^2 n \theta}{\sqrt{2}}   \: \Big(\ket {00}_{CC} \otimes\ket {T}_P\Big)+.........
\\& \: \: +\prod^{T-1}_{n=0}\frac{\cos^2 n \theta}{\sqrt{2}}   \: \Big(\ket {11}_{CC} \otimes\ket {-T}_{P}\Big).
\end{align}
Similarly by applying the walk operator on the initial state $\Phi_{\text{int2}}$ we get,

\begin{align}\notag
&\frac{1}{\sqrt{2}}(\ket {01}_{CC} +\ket {10}_{CC})\otimes \ket 0_P
\\\notag&\xRightarrow[\text{}]{1\text{st}}-\frac{1}{\sqrt{2}}(   \ket {01}_{CC} \otimes\ket {0}_{P}+ \ket {10}_{CC} \otimes\ket {0}_P),
\\ \notag &\xRightarrow[\text{}]{2\text{nd}}\frac{1}{\sqrt{2}}(  \ket {01}_{CC} \otimes\ket {0}_P+ \ket {10}_{CC} \otimes\ket {0}_P),
\\ \label{estate} &\xRightarrow[\text{}]{\text{Tth}}
(-1)^T\frac{1}{\sqrt{2}}\Big( \ket {01}_{CC} \otimes\ket {0}_P+ \ket {10}_{CC} \otimes\ket {0}_P\Big).
\end{align}

\subsection{Probability Distribution and Shannon Entropy}
We study the evolution of a walk driven by PDEC and compare it with the case of PIEC in order to demonstrate the role of position dependence. We carry out numerical simulations for the two initial states given in Eq. (\ref{int1}) and Eq. (\ref{int2}). For the initial state, $\Phi_{\text{int1}}$ the probability distribution for different rotation angles is shown in Fig. \ref{fig:f}. We observe the following properties:
\newline
(I) For $\theta=0$, the probability distribution of both types of walks, i.e. quantum walk with PDEC and the one with PIEC, show similar behavior. After T steps of both types of walks, the walker occupies positions $x=\pm T$ each with probability 0.5.
\newline
(II) For $\theta=\pi/2$, again the probability distribution of the quantum walk with PDEC shows similar behavior to the one with PIEC. The walker is localized at position $x=0$ for even number of steps. For odd values of T, the probability of the walker splits equally into two lattice sites $x=\pm 1$.
\newline
(III) For $\theta=\pi/4$, the probability distribution of the quantum walk with PDEC is periodic and bounded to the range of lattice sites $x=\pm 2$. For any value of T there are three possible configurations of the probability distribution: (a) The walker has probability 1 at position $x=0$ (localized), (b) the walker occupies positions $x=\pm 1$ each with probability $1/2$, (c) the walker occupies positions $x=0,\pm 1$ each with 1/4 probability, and $x=\pm 2$ both with 1/8 probability. For this rotation angle, the probability distribution of a walk with PIEC shows three-peaks-zone. By increasing the number of steps the probability increases in the initial position. The behavior of the probability distribution does not depend on the choice of the initial state in contrast to the case of PDEC.
\newline
(V) For $\theta=7\pi/45$, the probability distribution of the quantum walk with PDEC shows bounded behavior. Moreover, the walker has a higher probability at the initial position as compared to the other accessible lattice sites. In the case of PIEC, the number of occupied positions increases with the number of steps and the maximum probability is at positions $x=\pm T$.
\newline
(VI) For $\theta=\pi/5$, the quantum walk with PDEC shows that the probability increases at the initial position with the number of steps, i.e. the walker becomes more localized as compared to the walk with PIEC.
\newline
(VII) For $\theta=\pi/90$ the behavior of the quantum walk with PDEC shows maximum probability towards sides. Although there is a small probability near the initial position of the walker, as the number of steps increases the probability around the initial position also increases. For a quantum walk with PIEC, the probability is maximum at positions $x=\pm T$ because this rotation angle is closer to 0.
\newline
Generally, for the quantum walk with PDEC, we observe that the walker is more localized at the initial position (except for rotation angles closer to 0) as compared to quantum walk with PIEC. This statement holds true for other rotation angles like $\theta=\pi/12$ and $5\pi/18$ as well.
\par
For the second choice of the initial state ($\Phi_{\text{int2}}$) the probability distribution is plotted in Fig. \ref{fig:g} where our observations are as follows.
\newline
The stepwise evolution of a quantum walk with PDEC shows that the behavior of probability distribution does not depend on the choice of the rotation angle. The walker is completely localized at the initial position for any choice of the rotation angle and for any number of steps of the walk. This is because of the shift operator ($\hat{S}_{\text{EC}}$) which does not change the position of a walker prepared in this particular internal state. The localization here is different from the well known Anderson localization \cite{localization} which arises due to spatial disorder in a system. In our case, the localization is due to the shift operator $\hat{S}_{\text{EC}}$ which does not shift the walker in a particular internal state and thus remains trapped at the initial position in case of PDEC.
\par
The behavior of probability distribution in case of PIEC changes with rotation angles. Although the most probable position is $x=0$ for different rotation angles, still the probability distribution does not show complete localization as in the case of PDEC. This is also the case for $\theta=\pi/90$ where the probability is 0.8 at the initial position for PIEC.
\par
The Shannon entropy associated with position space of the quantum walk with PDEC is plotted as a function of T for the initial state $\Phi_{\text{int1}}$ (Fig. \ref{fig:h}). For a pure maximally entangled state the entropy associated with coin space is zero. For position space, we observe that the entropy is generally smaller for a walk with PDEC compared to PIEC (except $\theta=\pi/90$). This confirms the fact that PDEC brings localization to the walk as we have demonstrated on the basis of the probability distribution (see Fig. \ref{fig:f}). Additionally, for initial state $\Phi_{\text{int2}}$ the Shannon entropy is zero for any choice of rotation angle because of the localization of the walker.

\onecolumngrid

\begin{center}
\noindent
\begin{figure}[H]
\begin{minipage}{0.98\columnwidth}

\begin{tabular}{ccc}
      \addheight{\includegraphics[width=57mm]{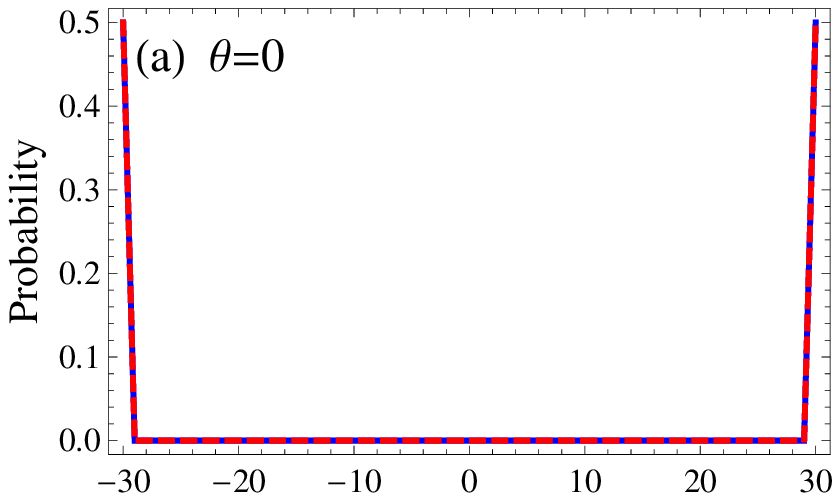}} &
      \addheight{\includegraphics[width=57mm]{pi2}} &
      \addheight{\includegraphics[width=57mm]{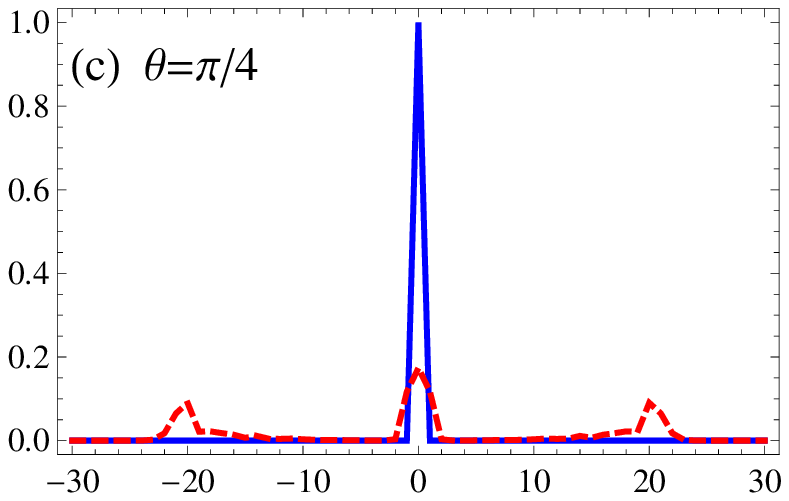}}\\
      \addheight{\includegraphics[width=57mm]{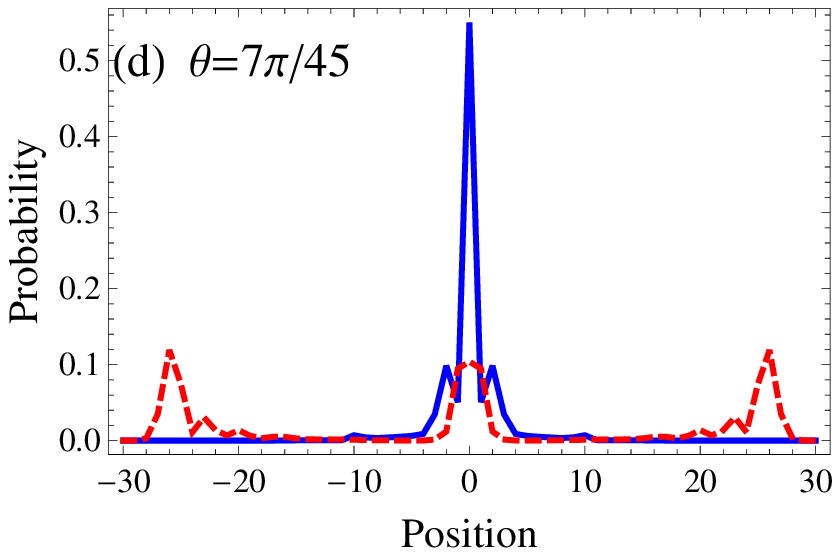}} &
     \addheight{\includegraphics[width=57mm]{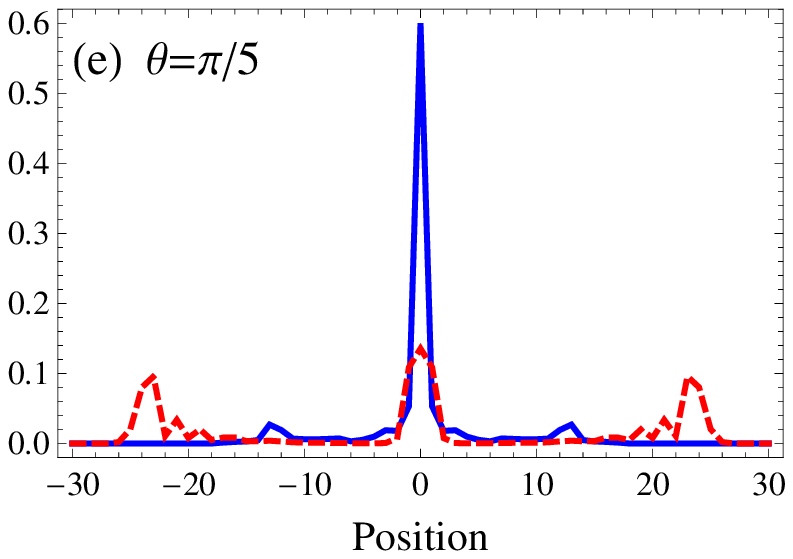}} &
     \addheight{\includegraphics[width=57mm]{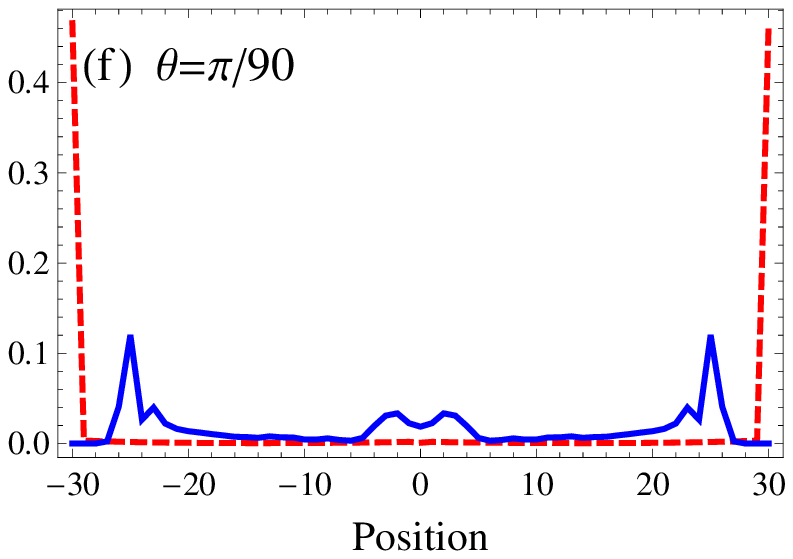}}\\
\end{tabular}
\caption{Probability vs Position for T=30 and initial state $\Phi_{\text{int1}}$. Solid blue line represents position-dependent entangled coins (PDEC) and dashed red line represents position-independent entangled coins (PIEC).}
\label{fig:f}
\end{minipage}
\end{figure}
\end{center}

\begin{center}
\noindent
\begin{figure}[H]
\begin{minipage}{0.98\columnwidth}

\begin{tabular}{ccc}
      \addheight{\includegraphics[width=57mm]{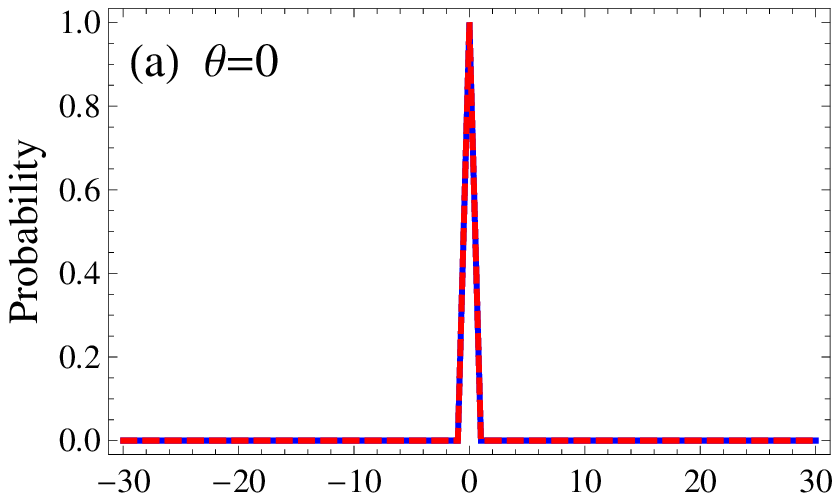}} &
       \addheight{\includegraphics[width=57mm]{pi2}} &
      \addheight{\includegraphics[width=57mm]{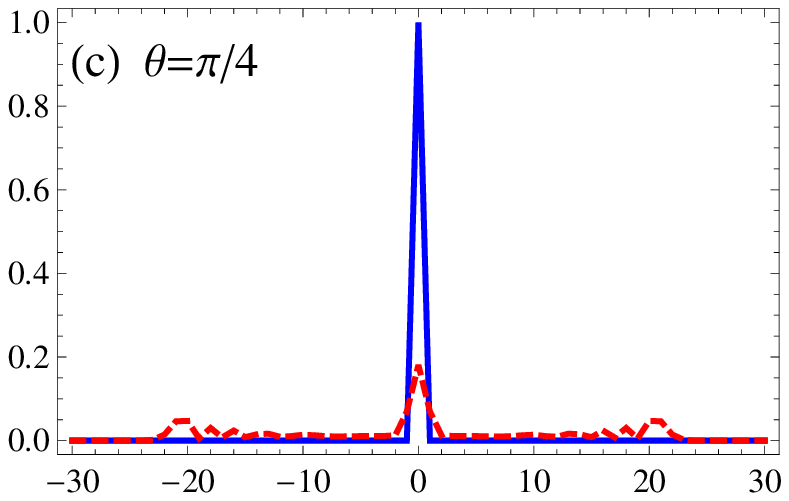}}\\
      \addheight{\includegraphics[width=57mm]{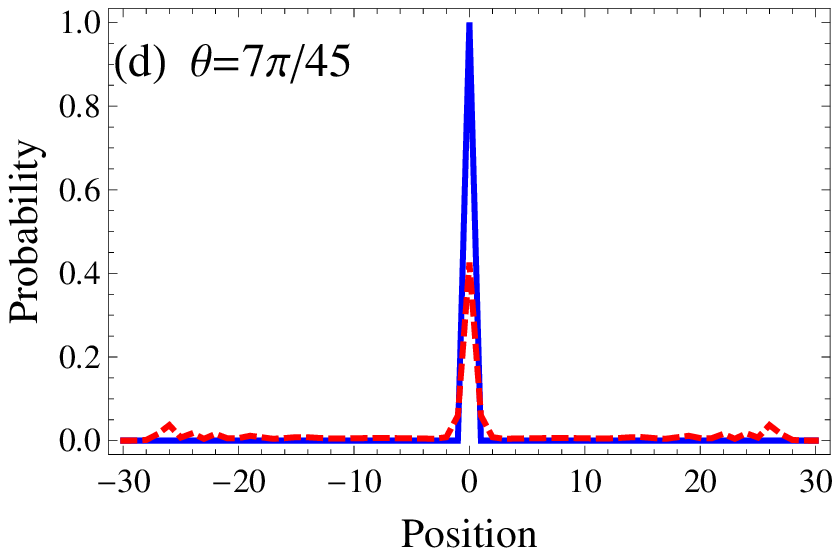}} &
     \addheight{\includegraphics[width=57mm]{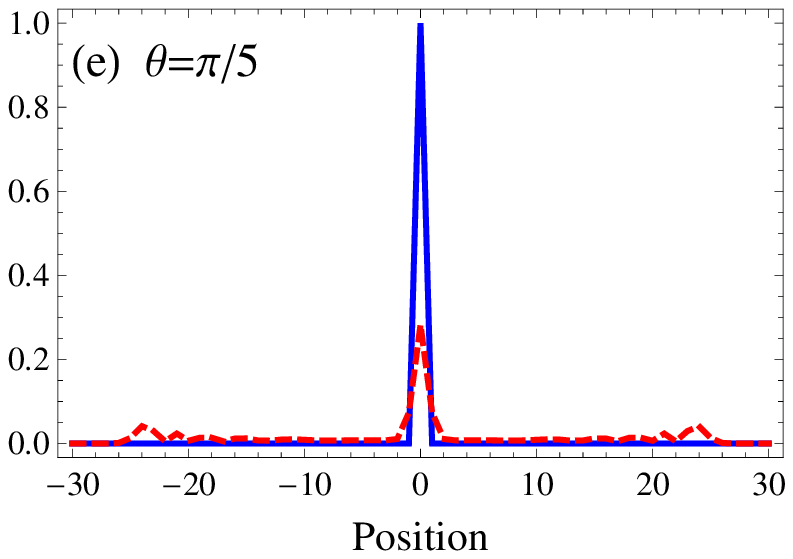}} &
     \addheight{\includegraphics[width=57mm]{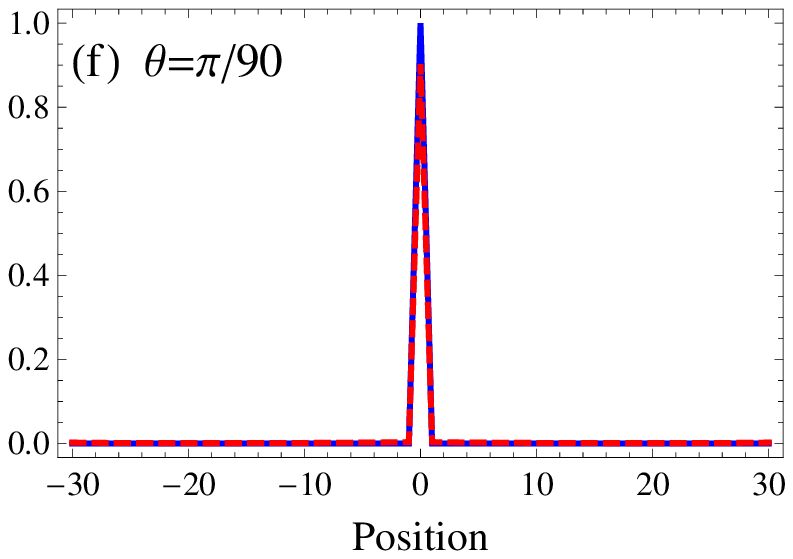}}\\
\end{tabular}
\caption{Probability vs Position for T=30  and initial state $\Phi_{\text{int2}}$. Solid blue line represents position-dependent entangled coins and dashed red line represents position-independent entangled coins.}
\label{fig:g}
\end{minipage}
\end{figure}
\end{center}

\begin{center}
\noindent
\begin{figure}[H]
\begin{minipage}{0.98\columnwidth}

\begin{tabular}{ccc}

      \addheight{\includegraphics[width=57mm]{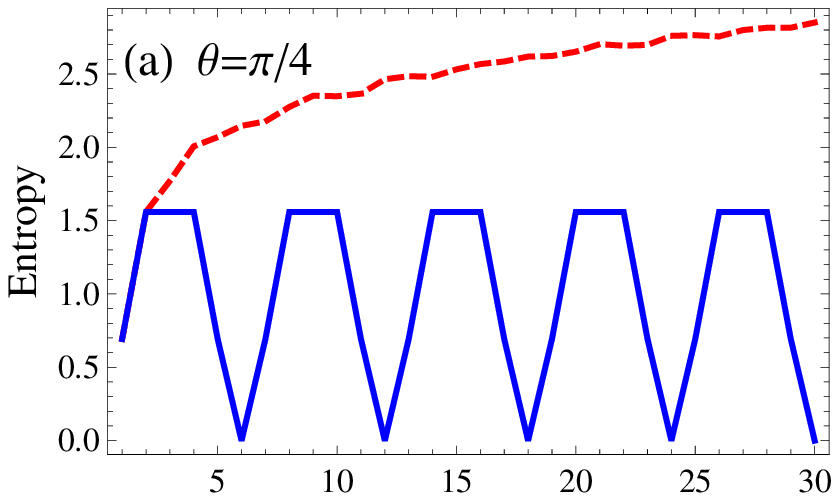}} &
     \addheight{\includegraphics[width=57mm]{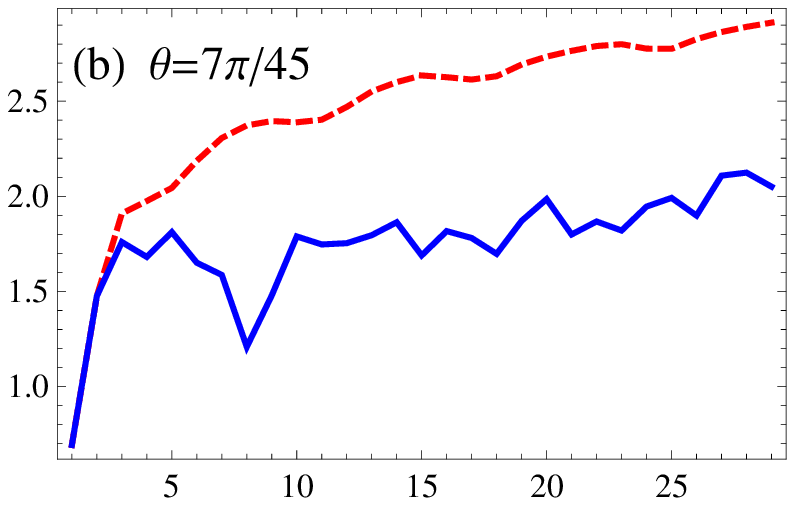}} &
     \addheight{\includegraphics[width=57mm]{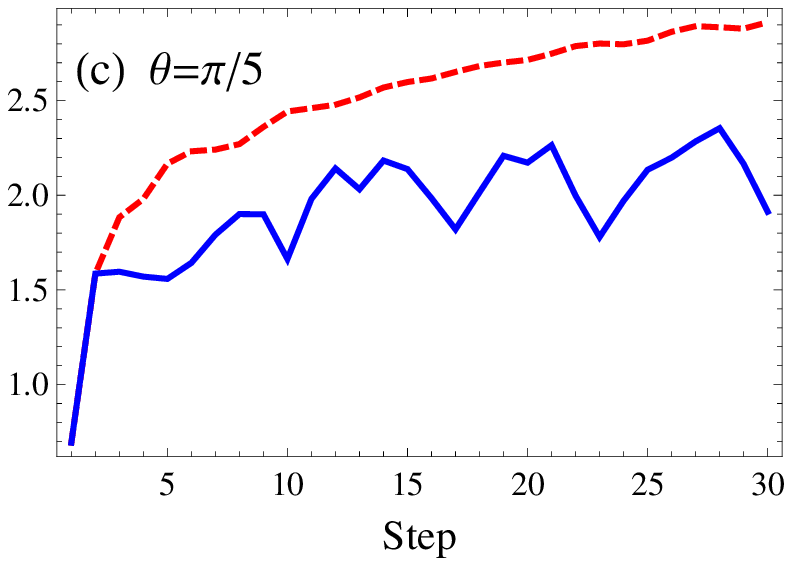}}\\
      \addheight{\includegraphics[width=57mm]{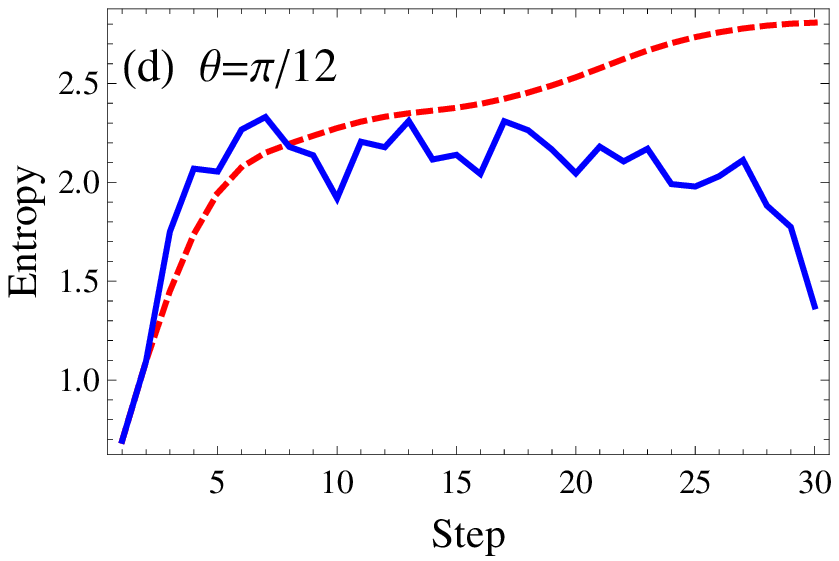}} &
      \addheight{\includegraphics[width=57mm]{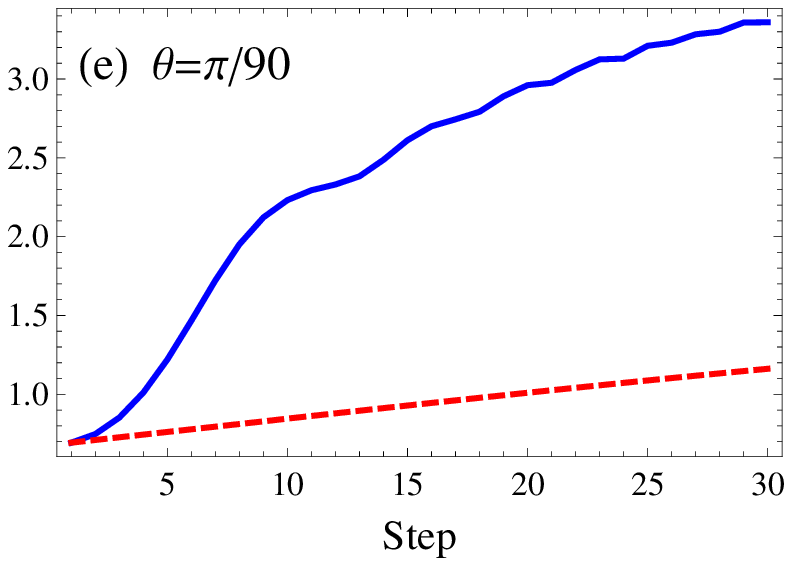}} \\
\end{tabular}
\caption{Entropy vs steps for T=30 and initial state $\Phi_{\text{int1}}$. Solid blue line represents the Shannon entropy of the quantum walk with position-dependent entangled coins (PDEC) and dashed red line represents the case of position-independent entangled coins (PIEC). Entropy is minimum for PDEC compared to PIEC for different rotation angles except for $\theta=\pi/90$.}
\label{fig:h}
\end{minipage}
\end{figure}
\end{center}

\twocolumngrid

\section{Conclusion}\label{conc}
We studied quantum walks in one dimension where the coin operator is characterized by a position-dependent rotation angle. We demonstrated that the walk shows diverse behavior for different rotation angles. Based on the probability distribution we classified the walk in different classes. This classification shows that PDC serves as a controlling tool to get different probability distributions where the controlling parameter is the rotation angle.
\par
Moreover, we studied the Shannon entropy of the quantum walk with PDC for different rotation angles and compared it with the case of PIC. We observed that the entropy in case of PDC is smaller as compared to the walk with PIC except for the bounded semi-classical-like and bounded quantum-like walk. This exception occurs because of the relatively more localized behavior of probability distribution in case of the walk with PIC. Generally, the entropy of the quantum walk with PIC increases with the number of steps irrespective of the rotation angle. In the case of a quantum walk with PDC, the behavior of entropy depends on the rotation angle.
\par
We further investigated 1D quantum walks with two identical PDEC. The study was conducted for two pure maximally entangled Bell states as the initial states of the quantum walk. We compared the probability distribution for PDEC to the case where the coins are position-independent. For the case of PDEC, we observed that the probability distribution of the walker becomes bounded to limited positions. The walker becomes more localized as compared to PIEC except for the smaller rotation angles, i.e. 0 and $\pi/90$. In the case of a specific initial state, the probability distribution showed that the walker becomes completely localized irrespective of the considered rotation angle. We also found that the Shannon entropy in the case of PDEC is smaller as compared to PIEC.

This study of PDC in a one-dimensional quantum walks can be generalized to study quantum walks with PDC in higher dimensions and with multi-particles. Moreover, the study of quantum walks with two PDEC can be generalized to more than two entangled qubits. The case of non-identical sub coins will be interesting to study. Quantum walks using coins with different values of entanglement will be worth to study. We will leave these matters for future work.
\section*{Acknowledgements}
Authors will like to thank S. Panahiyan  for his valuable suggestions!
\twocolumngrid

\end{document}